%% file: main.tex
\PassOptionsToPackage{table}{xcolor}
\documentclass[sigconf,screen]{acmart}

\copyrightyear{2026}
\acmYear{2026}
\setcopyright{cc}
\setcctype{by}
\acmConference[ASE '26]{Proceedings of the 41st IEEE/ACM International Conference on Automated Software Engineering}{October 12--16, 2026}{Munich, Germany}
\acmBooktitle{Proceedings of the 41st IEEE/ACM International Conference on Automated Software Engineering (ASE '26), October 12--16, 2026, Munich, Germany}
\acmDOI{10.1145/3832783.3834403}
\acmISBN{979-8-4007-2882-2/2026/10}

\usepackage[utf8]{inputenc}

\usepackage[utf8]{inputenc}
\usepackage[most]{tcolorbox}
\usepackage{xspace}
\usepackage{xspace}
\usepackage{pifont}
\usepackage{colortbl}
\usepackage{arydshln}
\usepackage{multirow}
\usepackage{enumitem}

\tcbuselibrary{listingsutf8}
\usepackage{listings}
\usepackage{xcolor}
\usepackage{tablefootnote}
\usepackage{wrapfig}
\usepackage{threeparttable}
\usepackage{arydshln}
\usepackage{subcaption}

\usepackage{warpcol}

\input{micro.tex}

\begin{document}

\title{Clarity Is Not Assumed: Understanding LLM-Based Code Generation under Ambiguous Requirements}

\author{Di Yang}
\orcid{0000-0003-3413-3138}
\affiliation{%
  \institution{East China Normal University}
  \city{}
  \country{}
}
\affiliation{%
  \institution{Shanghai Innovation Institute}
  \city{Shanghai}
  \country{China}
}
\email{diyang@stu.ecnu.edu.cn}

\author{Xinou Xie}
\orcid{0009-0007-3511-9457}
\affiliation{%
  \institution{East China Normal University}
  \city{Shanghai}
  \country{China}
}
\email{51275902162@stu.ecnu.edu.cn}

\author{Xiuwen Yang}
\orcid{0009-0007-9650-1705}
\affiliation{%
  \institution{East China Normal University}
  \city{Shanghai}
  \country{China}
}
\email{52295900006@stu.ecnu.edu.cn}

\author{Ming Hu}
\orcid{0000-0002-5058-4660}
\affiliation{%
  \institution{East China Normal University}
  \city{Shanghai}
  \country{China}
}
\email{mhu@sei.ecnu.edu.cn}

\author{Yihao Huang}
\orcid{0000-0002-5784-770X}
\affiliation{%
  \institution{East China Normal University}
  \city{Shanghai}
  \country{China}
}
\email{huangyihao@sei.ecnu.edu.cn}

\author{Yueling Zhang}
\orcid{0000-0002-2542-2619}
\affiliation{%
  \institution{East China Normal University}
  \city{Shanghai}
  \country{China}
}
\email{ylzhang@sei.ecnu.edu.cn}

\author{Weikai Miao}
\orcid{0000-0002-2308-0926}
\affiliation{%
  \institution{East China Normal University}
  \city{Shanghai}
  \country{China}
}
\email{wkmiao@sei.ecnu.edu.cn}

\author{Ting Su}
\orcid{0000-0003-1628-9796}
\affiliation{%
  \institution{East China Normal University}
  \city{Shanghai}
  \country{China}
}
\email{tsu@sei.ecnu.edu.cn}

\author{Chengcheng Wan}
\orcid{0000-0001-9162-9688}
\authornote{Chengcheng Wan is the corresponding author.}
\affiliation{%
  \institution{East China Normal University}
  \city{}
  \country{}
}
\affiliation{%
  \institution{Shanghai Innovation Institute}
  \city{Shanghai}
  \country{China}
}
\email{ccwan@sei.ecnu.edu.cn}

\author{Geguang Pu}
\orcid{0000-0001-9750-8334}
\affiliation{%
  \institution{East China Normal University}
  \city{Shanghai}
  \country{China}
}
\email{ggpu@sei.ecnu.edu.cn}

\renewcommand{\shortauthors}{D. Yang, X. Xie, X. Yang, M. Hu, Y. Huang, Y. Zhang, W. Miao, T. Su, C. Wan, and G. Pu}

\begin{abstract}

Software requirement ambiguity is ubiquitous in real-world development, stemming from the inherent imprecision of natural language and the varying interpretations of stakeholders. While Large Language Models (LLMs) have demonstrated impressive capabilities in generating code from precise specifications, such ambiguity poses a significant obstacle to reliable automated code generation. Existing benchmarks typically assume clear and unambiguous requirements, leaving an empirical gap in understanding how LLMs behave when faced with the inherent uncertainty of real-world software requirements.

In this paper, we introduce \textbf{\bench}, a code generation benchmark specifically designed with ambiguous requirements. It comprises 1,304 function-level tasks covering four distinct types of ambiguity: \textit{lexical}, \textit{syntactic}, \textit{semantic}, and \textit{vagueness}. Leveraging this dataset, we conduct the first systematic empirical study to evaluate the impact of requirement ambiguity on LLM-based code generation. Our results demonstrate that ambiguity consistently degrades the performance of all evaluated LLMs, with the most pronounced negative effects observed in highly advanced models. Furthermore, we observe that LLMs frequently produce functionally divergent implementations for the same ambiguous requirement and lack the capability to identify or resolve such ambiguity autonomously. These findings reveal a significant performance gap between clear and ambiguous requirements, underscoring the urgent need for ambiguity-aware techniques in the next generation of automated software engineering tools. The Orchid benchmark is publicly available at \url{https://huggingface.co/datasets/SII-YDD/Orchid}.

\end{abstract}


\maketitle

\section{Introduction}

\input{sections/1-introduction.tex}

\section{Background}
\input{sections/2-background.tex}

\section{Requirement Ambiguity}
\input{sections/3-definition_and_taxonomy_of_code_generation_ambiguity.tex}

\section{\bench Construction}
\input{sections/4-benchmark_design_and_dataset.tex}

\section{Benchmarking Analysis}
\input{sections/5-the_impact_of_requirement_ambiguity.tex}

\section{Learned Lessons}
\input{sections/6-learned_lessons.tex}

\section{Related Work}
\input{sections/7-related_work.tex}

\section{Limitations}
\input{sections/8-limitations.tex}

\section{Conclusion}

\input{sections/9-conclusion.tex}

\begin{acks}
This work was supported by the National Natural Science Foundation of China (Grant Nos. 62402183, 92582108, 62572192, 62372181). It is also supported by the Shanghai Special Program for Promoting High-Quality Industrial Development (Project No. 250401) and the Fundamental Research Funds for the Central Universities.
\end{acks}

\section{Data Availability Statement}
The artifact for this study includes the \bench benchmark (comprising 1,304 tasks and 5,216 requirements), the multi-agent ambiguity construction pipeline, and the raw experimental outputs from all evaluated LLMs. It is available on Figshare~\cite{orchid2026}.

\bibliographystyle{ACM-Reference-Format}
\bibliography{citation.bib}

\end{document}

%% file: micro.tex
\newcommand{\bench}{Orchid\xspace}

\definecolor{todocolor}{rgb}{0.9,0.1,0.1}

\newcommand{\eg}{\hbox{\emph{e.g.}}\xspace}
\newcommand{\ie}{\hbox{\emph{i.e.}}\xspace}
\newcommand{\st}{\hbox{\emph{s.t.}}\xspace}

\usepackage{pythonhighlight}

\definecolor{codegreen}{rgb}{0,0.6,0}
\definecolor{codegray}{rgb}{0.5,0.5,0.5}
\definecolor{codepurple}{rgb}{0.58,0,0.82}
\definecolor{backcolour}{rgb}{0.97,0.97,0.95}
\definecolor{forestgreen}{rgb}{0.28,0.62,0.37}
\definecolor{skyblue}{RGB}{174,205,225}

\lstdefinestyle{mystyle}{
    backgroundcolor=\color{backcolour},   
    keywordstyle=\color{codepurple},
    numberstyle=\tiny\color{codegray},
    stringstyle=\color{blue},
    basicstyle=\ttfamily\footnotesize,
    breakatwhitespace=false,         
    breaklines=true,                 
    captionpos=b,                    
    keepspaces=true,                 
    numbers=left,                    
    numbersep=5pt,                  
    showspaces=false,                
    showstringspaces=false,
    showtabs=false,                  
    tabsize=4,
}

\lstdefinestyle{mypython}{
  language=Python,
  basicstyle=\ttfamily\small,
  keywordstyle=\color{blue},
  commentstyle=\color{black},
  stringstyle=\color{orange},
  showstringspaces=false,
  breaklines=true,
  escapeinside={(*@}{@*)},
  backgroundcolor=\color{gray!10},
  columns=flexible
}

\newtcolorbox{ambiguityexample}[2][]{%
  enhanced,
  colback=gray!10,
  colframe=black!80,
  arc=2mm,
  boxrule=0.5pt,
  title={#2},
  listing only,
  listing options={style=mypython},
  #1,
  before upper={\vspace{-3mm}},
  after upper={\vspace{-3mm}},
}

\newcommand{\boxmargin}{1mm}
\newtcolorbox{myboxc}{
    colback=gray!15!white,
    arc = 0pt, outer arc = 0pt,
    boxsep=0pt, left = 3pt, right = 0pt, top = 0pt, bottom = 0pt, 
    leftrule=3pt, bottomrule=0pt,toprule=0pt, rightrule=0pt,
    left = \boxmargin, right = \boxmargin, top = \boxmargin, bottom = \boxmargin
}


%% file: sections/1-introduction.tex
As emphasized by \citet{brooks1987essence}, determining precisely what to build remains one of the most challenging aspects of software development. This challenge lies at the core of Requirements Engineering (RE), which forms the foundation of the software development life cycle. In practice, requirements are predominantly documented in natural language, which is inherently imprecise and prone to ambiguity~\cite{berry2004ambiguity,bano2015addressing,ezzini2021using}. Such ambiguity, where a single description corresponds to multiple conflicting interpretations, is prevalent in real-world development due to limited communication and the varying expertise of stakeholders~\cite{gervasi2005reasoning,fischbach2021practitioners}. 

\begin{figure}
  \centering
  \includegraphics[width=0.95\linewidth]{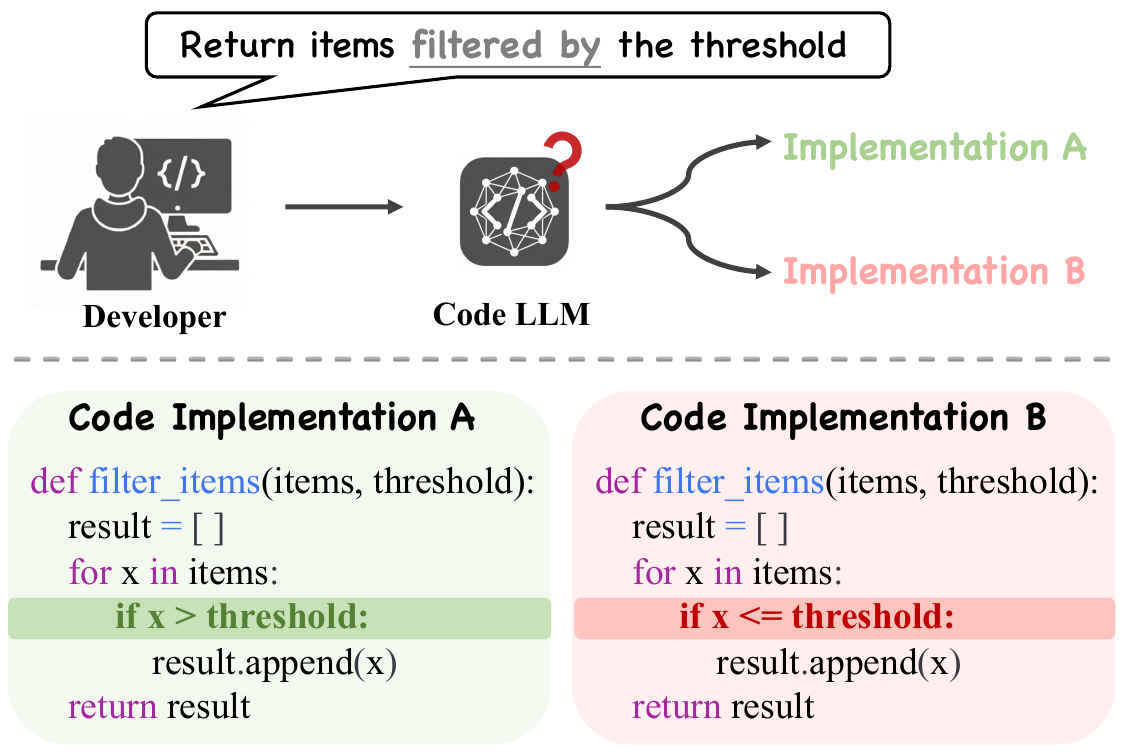}
  \caption{An ambiguous requirement results in LLMs generating functionally distinct code snippets. Here, \textit{"filtered by"} is implemented in two different ways: (\textcolor[RGB]{162,200,137}{A}) retaining items above the threshold; and (\textcolor[RGB]{234,153,153}{B}) keeping only those below it.}
  \label{fig:ambiguity_motivation}
\end{figure}

While human developers can mitigate these uncertainties through iterative clarification, Large Language Model (LLM) based code generation faces a critical conflict. LLM-based solutions are typically forced into determinism, which collapses inherent linguistic uncertainty into a single, executable implementation without explicit clarification. As illustrated in Figure~\ref{fig:ambiguity_motivation}, an ambiguous requirement using the phrase \textit{``filtered by''} results in LLMs generating functionally distinct code snippets. The ambiguity leads to two divergent interpretations: \textcolor[RGB]{162,200,137}{\textbf{Implementation A}} retaining items above the threshold (\texttt{x > threshold}); \textcolor[RGB]{234,153,153}{\textbf{Implementation B}} keeping only those below it (\texttt{x <= threshold}). This forced determinism compels models to make unwarranted assumptions, posing a significant barrier to reliable code generation.

Despite the rapid progress in LLM-based code generation, widely used function-level benchmarks generally assume that each natural-language requirement has a single intended interpretation, largely excluding linguistic uncertainty from evaluation~\cite{liu2023your,zhuo2024bigcodebench}. In parallel, the requirements engineering (RE) community has studied ambiguity from a human-centric perspective~\cite{ferrari2019nlp,gentili2023characterizing}. However, how code LLMs behave when confronted with inherently ambiguous requirements remains underexplored, limiting our understanding of their reliability in realistic software development settings. This gap stems from a fundamental limitation in existing evaluation paradigms: ambiguity is neither explicitly modeled nor systematically controlled. As a result, prior studies are unable to disentangle whether failures arise from model deficiencies or from inherent uncertainty in the input specification.

To address this gap, we introduce \textbf{\bench}, a function-level benchmark designed to evaluate the impact of requirement ambiguity on code generation. \bench consists of 1,304 tasks and 5,216 ambiguous requirement variants, covering four types of ambiguity: \textit{lexical}, \textit{syntactic}, \textit{semantic}, and \textit{vagueness}. These types represent different sources of interpretive uncertainty, allowing us to assess how varying linguistic factors affect model behavior. The benchmark is constructed using a semi-automated pipeline that follows a general, reusable method for ambiguity data construction. This method is based on a multi-agent framework designed for ambiguity injection, ensuring both scalability and high-quality output. The process begins with the Injection Agent, which introduces ambiguity into each clean functional requirement based on predefined ambiguity types. After ambiguity is injected, the Judge Agent evaluates the variants to ensure that the injected ambiguity is contextually plausible and retains the original functional intent. Finally, the Explain Agent provides concise explanations of the plausible interpretations for each ambiguous requirement, clarifying the effects of the ambiguity. To ensure the quality of the manually curated core, experts conducted item-level validation of all 1,312 variants in Orchid-HEval and Orchid-BCB, requiring over 246 person-hours.

\begin{table*}
\footnotesize
\centering
\caption{Code Generation Benchmarks.}
\begin{threeparttable}
\resizebox{\linewidth}{!}{%
\rowcolors{2}{gray!10}{white} 
\begin{tabular}{ l c c l c c c}
\toprule
\textbf{Benchmark} & \textbf{Time} & \textbf{Granularity} & \textbf{\#Tasks} & \textbf{Construction} & \textbf{Language} & \textbf{Req. Characteristics$^2$} \\
\midrule
CoNaLa~\cite{yin2018learning}                & 2018 & Statement-level    & 500   & Semi-automated  & Python         & Code-aligned \\
HumanEval~\cite{chen2021evaluating}          & 2021 & Function-level     & 164   & Manual          & Python         & Concise \\
MBPP~\cite{austin2021program}                & 2021 & Function-level     & 974   & Manual          & Python         & Concise \\
MBXP~\cite{athiwaratkun2022multi}           & 2022 & Function-level     & 974   & Manual          & Multi-language       & Diverse \\
Multi-HumanEval~\cite{athiwaratkun2022multi} & 2022 & Function-level     & 164   & Manual          & Multi-language       & Concise \\
DS-1000~\cite{lai2023ds}                     & 2022 & Statement-level    & 1000  & Semi-automated  & Python         & Code-aligned \\
HumanEval+~\cite{liu2023your}                & 2023 & Function-level     & 164   & Automated          & Python         & Concise \\
SWE-Bench~\cite{jimenez2023swe}              & 2023 & Pull Request-level & 2294  & Semi-automated  & Python         & Contextual \\
CRUXEval~\cite{gu2024cruxeval}               & 2024 & Function-level     & 800   & Semi-automated  & Python         & Multi-step \\
ClassEval~\cite{du2024evaluating}            & 2024 & Class-level        & 100   & Manual          & Python         & Structured \\
Safim~\cite{gong2024evaluation}              & 2024 & Statement-level    & 17720 & Semi-automated  & Multi-language       & Code-aligned \\
CanItEdit~\cite{cassano2023can}              & 2024 & Function-level     & 105   & Manual          & Python         & Instructional \\
LiveCodeBench~\cite{jain2024livecodebench}  & 2024 & Function-level     & 1055\tnote{1} & Semi-automated & Python & Concise \\
BigCodeBench~\cite{zhuo2024bigcodebench}     & 2024 & Function-level     & 1140  & Semi-automated  & Python         & Concise  \\
\hdashline
\rowcolor{white}
\textbf{\bench (Ours)}                       & \textbf{2026} & \textbf{Function-level} & \textbf{1304} & \textbf{Semi-automated} & \textbf{Python} & \textbf{Ambiguity} \\
\bottomrule
\end{tabular}
}
\begin{tablenotes}
\scriptsize
\item[1] As of release\_v6 (Apr 2025). 
\item[2] Characteristics of code generation requirements.
\end{tablenotes}
\end{threeparttable}
\label{tab:benchmark-comparison}
\end{table*}
To delve into the challenges posed by requirement ambiguity, we conduct a comprehensive empirical study using \bench, focusing on three critical dimensions of LLM behavior. First, we investigate the performance impact of ambiguity (\textbf{RQ1: How does ambiguity impact LLM performance?}). Our analysis reveals a pervasive and substantial degradation in generation quality across all evaluated models. Even state-of-the-art models, such as GPT-4, exhibit a performance drop exceeding 30\% when confronted with ambiguous specifications, suggesting that current benchmarks significantly overestimate the effectiveness of LLMs in real-world, "noisy" software engineering scenarios. Second, we assess the functional consistency of LLMs under uncertainty (\textbf{RQ2: How consistent are LLMs in generating functional code under ambiguity?}). Beyond mere correctness, we find that ambiguity undermines the reliability of code generation by triggering functional divergence. Models frequently produce multiple, mutually incompatible implementations for the same ambiguous prompt across different trials. This lack of determinism indicates that LLMs struggle to maintain a stable internal representation when requirements are not strictly bounded. Third, we evaluate the models' intrinsic capability to recognize and resolve ambiguity (\textbf{RQ3: How well can LLMs recognize and resolve ambiguities?}). While LLMs demonstrate a surprisingly high recall in flagging potential ambiguities, they suffer from overprediction and uncertainty. More importantly, they consistently fail to precisely localize the source of ambiguity or provide valid resolutions, highlighting a fundamental gap between detecting a problem and understanding its logic. 

These findings underscore that ambiguity is not merely a data-level noise but a structural bottleneck that compromises the trustworthiness of automated code generation. In summary, this paper makes the following contributions:
\begin{itemize}[leftmargin=*, noitemsep]
    \item \textbf{Ambiguity Benchmark:} 
    We constructed \bench, a benchmark of 1,304 tasks and 5,216 requirements 
    spanning four ambiguity types, enabling evaluation of LLMs 
    under ambiguity.
    \item \textbf{Empirical Impact Study:} 
    We systematically quantify the impact of ambiguity on LLM performance and functional consistency across multiple models.
    \item \textbf{LLM Behavior Analysis:} 
    We characterize how LLMs handle ambiguous requirements, focusing on 
    their ability to recognize ambiguity and their limitations in 
    localizing and resolving it.
\end{itemize}

%% file: sections/2-background.tex
\begin{table*}
\centering
\caption{Types of Ambiguities in Function-Level Code Generation~\cite{shah2015resolving}.}
\label{tab:ambiguity_types}
\small
\resizebox{\linewidth}{!}{%
\begin{tabular}{p{1.5cm} p{13.5cm}}
\toprule
\makebox[1.5cm][l]{\textbf{Types}} & 
\multicolumn{1}{c}{\textbf{Definition}} \\
\midrule
\makebox[1.5cm][l]{\textbf{Lexical}} & 
A word in the requirement has multiple distinct meanings, causing different interpretations. \\
\makebox[1.5cm][l]{\textbf{Syntactic}} & 
A requirement's grammatical structure has multiple distinct meanings, causing different interpretations. \\
\makebox[1.5cm][l]{\textbf{Semantic}} & 
A phrase or sentence in the requirement has multiple distinct meanings, causing different interpretations. \\
\makebox[1.5cm][l]{\textbf{Vagueness}} & 
A requirement that omits necessary details has multiple distinct meanings, causing different interpretations. \\
\bottomrule
\end{tabular}
}
\end{table*}

\subsection{LLM-Based Code Generation}

Large language models (LLMs) have significantly improved code generation from natural language requirements, enabling the production of syntactically correct and semantically meaningful programs~\cite{nijkamp2022codegen,chen2021evaluating,wang2021codet5}. These models achieve strong performance when requirements are clearly specified, where the intended functionality can be directly inferred from the input.

Code generation relies on the model's interpretation of natural language. When a requirement admits multiple plausible interpretations, the generation process becomes underdetermined. In such cases, the model must resolve ambiguity implicitly and produce a single implementation without external clarification. Different interpretations can therefore lead to functionally different outputs. Despite the importance of this problem, the impact of requirement ambiguity on code generation remains insufficiently understood.

\subsection{Code Generation Benchmarks}

Code generation benchmarks provide standardized datasets for evaluating a model's ability to translate natural language requirements into executable programs. Existing benchmarks cover different levels of granularity, including statement-level tasks that focus on syntactic correctness~\cite{yin2018learning,lai2023ds,gong2024evaluation}, function-level tasks that evaluate semantic correctness and reasoning~\cite{chen2021evaluating,austin2021program,athiwaratkun2022multi,liu2023your,jain2024livecodebench,zhuo2024bigcodebench,cassano2023can,gu2024cruxeval}, and higher-level tasks that involve classes or repositories~\cite{du2024evaluating,jimenez2023swe}.

As shown in Table~\ref{tab:benchmark-comparison}, widely used benchmarks such as HumanEval~\cite{chen2021evaluating}, MBPP~\cite{austin2021program}, and BigCodeBench~\cite{zhuo2024bigcodebench} adopt concise and well-defined requirements. These datasets are constructed through manual curation or semi-automated pipelines with human validation, which ensures clarity and consistency in task descriptions.

Most existing benchmarks share a common assumption that each requirement has a single intended interpretation. This assumption simplifies evaluation and enables consistent measurement of functional correctness. However, real-world requirements often contain ambiguity, where multiple interpretations are possible. As a result, current benchmarks mainly evaluate model performance under idealized conditions and do not capture model behavior under ambiguous requirements.

\subsection{Ambiguity in Software Requirements}

Ambiguity in software requirements arises when a specification allows multiple valid interpretations~\cite{berry2004ambiguity,shah2015resolving}. It can result from insufficient information, such as under-specification or vagueness, as well as linguistic factors including lexical, semantic, and syntactic ambiguity. Prior work in requirements engineering has studied ambiguity detection and mitigation, with a primary focus on supporting human developers~\cite{ferrari2019nlp,gentili2023characterizing}.

In LLM-based code generation, ambiguity introduces a fundamental challenge. Human developers can iteratively refine and clarify requirements~\cite{fischbach2021practitioners}, while LLMs are typically required to produce a single output given the input. This mismatch leads to inconsistent or incorrect implementations when different interpretations are possible. In addition, different runs or different models resolve ambiguity in different ways, resulting in functional divergence.

%% file: sections/3-definition_and_taxonomy_of_code_generation_ambiguity.tex
\label{sec:ambiguity-types}

As shown in Table~\ref{tab:ambiguity_types}, our paper focuses on function-level requirements and considers four types of ambiguity --- lexical, semantic, syntactic, and vagueness, following a study of natural language software requirements~\cite{shah2015resolving}. We omit pragmatic ambiguity, as it involves implied intentions or assumptions that rarely appear in function-level requirements. We also omit language errors and generality problems, as they are less relevant to code generation.

Not all ambiguity of requirements could lead to confusion when generating the code. We regard the ambiguity as taking effect if it could be interpreted in multiple plausible ways, each corresponding to a functionally different implementation. Formally, an NL requirement $R$ is ambiguous for code generation only when 
\begin{align}
&\exists I, I' \in \mathbb{I}(R), \st\,\, \mathrm{F}_{I} \neq \mathrm{F}_{I'}, \\
&\mathrm{F}_{I} \neq \mathrm{F}_{I'} \iff \exists x, \st\,\, \mathrm{F}_{I}(x)\neq\mathrm{F}_{I'}(x)
\end{align}
where $\mathbb{I}(R)$ is the set of all plausible interpretations of $R$, $\mathrm{F}_{I}$ is the functionality of interpretation $I$, and $\mathrm{F}_{I}(x)$ is the expected output when given an input $x$.

\subsection{Lexical Ambiguity} 
\begin{figure}[h]
    \centering
    \begin{subfigure}{\columnwidth}
        \centering
\begin{lstlisting}[language=Python, escapechar=\%]
from typing import List
def filter_by_substring(strings: List[str], substring: str) -> List[str]:
    Filter an input list of strings only for ones that 
    contain given %\textbf{pattern.}%
\end{lstlisting}
        \caption{Lexical ambiguity in Orchid-HEval task \#7.}
        \label{fig:lexical_ambiguity_example}
    \end{subfigure}

    \begin{subfigure}{\columnwidth}
        \centering
\begin{lstlisting}[language=Python, escapechar=\%]
%\textcolor[RGB]{162,200,137}{\textbf{\# Interpretation A}}%
def filter_by_substring(strings: List[str], substring: str) -> List[str]:
    return [s for s in strings if substring in s]
\end{lstlisting}
\begin{lstlisting}[language=Python, escapechar=\%]
%\textcolor[RGB]{234,153,153}{\textbf{\# Interpretation B}}%
def filter_by_substring(strings: List[str], substring: str) -> List[str]:
    regex = re.compile(pattern)
    return [s for s in strings if regex.search(s)]
\end{lstlisting}
        \caption{Divergent implementations.}
        \label{fig:lexical_ambiguity_implementations}
    \end{subfigure}
    \caption{A lexical ambiguity example from Orchid-HEval task \#7, where 'pattern' is interpreted in multiple ways.}
    \label{fig:overall_analysis}
\end{figure}

Lexical ambiguity appears when a word in the requirement has multiple meanings that are seemingly feasible in its context. It often arises from polysemous words with related senses or from vague terms with an undefined scope. In software requirements, this ambiguity is harmful because the interpretation of a single word can even alter the entire implementation. 

Figure~\ref{fig:lexical_ambiguity_example} illustrates an example of lexical ambiguity in the requirement. The requirement uses the term ``pattern,'' which can be interpreted either as a literal substring or as a more general pattern such as a regular expression, leading to multiple possible implementations. Specifically, this ambiguity may lead to two different implementations. Implementation A (Figure~\ref{fig:lexical_ambiguity_implementations}) checks if each string contains the input as a literal substring and returns those that do. Implementation B (Figure~\ref{fig:lexical_ambiguity_implementations}) treats the input as a regular expression pattern, using regex matching to return all strings that satisfy the pattern.  For test case \texttt{strings = ["aaa", "aa", "a", "b"]}, \texttt{substring = "a*"}, implementation A returns an empty list \texttt{[]}, while implementation B returns \texttt{["aaa", "aa", "a"]}.

\subsection{Syntactic Ambiguity} 
\begin{figure}[h]
    \centering
    \begin{subfigure}{\columnwidth}
        \centering
\begin{lstlisting}[language=Python, escapechar=\%]
def unique(l: list):
    %\textbf{Sort the list}% and return the unique elements in %\textbf{it.}%
\end{lstlisting}
        \caption{Syntactic ambiguity in Orchid-HEval task \#34.}
        \label{fig:syntactic_ambiguity_example}
    \end{subfigure}
    \begin{subfigure}{\columnwidth}
        \centering
\begin{lstlisting}[language=Python, escapechar=\%]
%\textcolor[RGB]{162,200,137}{\textbf{\# Interpretation A}}%
def unique(l: list):
    return sorted(set(l))
\end{lstlisting}
\begin{lstlisting}[language=Python, escapechar=\%]
%\textcolor[RGB]{234,153,153}{\textbf{\# Interpretation B}}%
def unique(l: list):
    sorted_l = sorted(l)
    return list(set(l))
\end{lstlisting}
        \caption{Divergent implementations.}
        \label{fig:syntactic_ambiguity_implementations}
    \end{subfigure}
    \caption{A syntactic ambiguity example from Orchid-HEval task \#34, where the pronoun ``it'' has an unclear reference.}
    \label{fig:syntactic_overall_analysis}
\end{figure}

Syntactic ambiguity occurs when the grammatical structure of a requirement sentence allows multiple valid interpretations, creating uncertainty in how its components are organized and related. Resolving syntactic ambiguity through context alone is often challenging, particularly for complex or lengthy sentences, where multiple parses may remain plausible despite rich contextual cues. Furthermore, syntactic ambiguity can fundamentally change the requirements' logic and behavior by affecting interpretations of action order, conditional scopes, and other critical aspects, thus causing risks to the correct implementation.

Figure~\ref{fig:syntactic_ambiguity_example} illustrates an example of syntactic ambiguity. The requirement involves returning the unique elements of a list, but the pronoun ``it'' can refer either to the sorted list or to the original unsorted list. This syntactic ambiguity results in two implementations. Implementation A (Figure~\ref{fig:syntactic_ambiguity_implementations}) returns the unique elements in sorted order by directly applying a set to the list and then sorting the result. Implementation B (Figure~\ref{fig:syntactic_ambiguity_implementations}) first sorts the list (though this sorted list is unused), then returns the unique elements of the original list without preserving any particular order. For test case \texttt{l = [4, 3, 3, 1, 2, 5, 1]}, implementation A returns \texttt{[1, 2, 3, 4, 5]}, while implementation B returns \texttt{[4, 3, 1, 2, 5]}.

\subsection{Semantic Ambiguity}  
\begin{figure}[h]
    \centering
    \begin{subfigure}{\columnwidth}
        \centering
\begin{lstlisting}[language=Python, escapechar=\%]
def remove_duplicates(numbers: List[int]) -> List[int]:
    From a list of integers, remove %\textbf{duplicate occurrences,}%
    preserving the initial order of the remaining elements.%
\end{lstlisting}
        \caption{Semantic ambiguity in Orchid-HEval task \#26.}
        \label{fig:semantic_ambiguity_example}
    \end{subfigure}

    \begin{subfigure}{\columnwidth}
        \centering
\begin{lstlisting}[language=Python, escapechar=\%]
%\textcolor[RGB]{162,200,137}{\textbf{\# Interpretation A}}%
def remove_duplicates(numbers: List[int]) -> List[int]:
    result = []
    for x in numbers:
        if numbers.count(x) == 1:
            result.append(x)
    return result
\end{lstlisting}

\begin{lstlisting}[language=Python, escapechar=\%]
%\textcolor[RGB]{234,153,153}{\textbf{\# Interpretation B}}%
def remove_duplicates(numbers: List[int]) -> List[int]:
    seen = set()
    result = []
    for x in numbers:
        if x not in seen:
            seen.add(x)
            result.append(x)
    return result
\end{lstlisting}
        \caption{Divergent implementations.}
        \label{fig:semantic_ambiguity_implementations}
    \end{subfigure}
    \caption{A semantic ambiguity from Orchid-HEval task \#26, where 'duplicate occurrences' is interpreted differently.}
    \label{fig:semantic_overall_analysis}
\end{figure}

Semantic ambiguity occurs when a phrase or sentence contains expressions that allow multiple plausible interpretations within the context of the requirement. This means that the same part of the requirement can be reasonably understood in different ways, potentially leading to varied implementations. Compared to lexical ambiguity, resolving this ambiguity requires broader contextual information and deeper reasoning.

Figure~\ref{fig:semantic_ambiguity_example} illustrates an example of semantic ambiguity. The requirement instructs the removal of duplicate occurrences in a list, which allows for two plausible interpretations:  Implementation A (Figure~\ref{fig:semantic_ambiguity_implementations}) filters out all repeated elements, returning only those with a single occurrence. Implementation B (Figure~\ref{fig:semantic_ambiguity_implementations}) preserves the first occurrence of each integer, removing subsequent duplicates while maintaining the original order. For test case \texttt{numbers = [1, 2, 2, 3, 4, 4, 5]}, implementation A returns \texttt{[1, 3, 5]}, while implementation B returns \texttt{[1, 2, 3, 4, 5]}.

\subsection{Vagueness Ambiguity}  
\begin{figure}[h]
    \centering
    \begin{subfigure}{\columnwidth}
        \centering
\begin{lstlisting}[language=Python, escapechar=@]
def digits(n: int) -> int:
    Given a positive integer n, return the product of the
    @\textbf{digits.}@
\end{lstlisting}
        \caption{Vagueness ambiguity in \bench-HEval task \#131.}
        \label{fig:vagueness_ambiguity_example}
    \end{subfigure}
    \begin{subfigure}{\columnwidth}
        \centering
\begin{lstlisting}[language=Python, escapechar=@]
@\textcolor[RGB]{162,200,137}{\textbf{\# Interpretation A}}@
def digits(n):
    odd_digits = []
    for d in str(n):
        if int(d) % 2 != 0: odd_digits.append(int(d))
    return math.prod(odd_digits) if odd_digits else 0
\end{lstlisting}
\begin{lstlisting}[language=Python, escapechar=@]
@\textcolor[RGB]{234,153,153}{\textbf{\# Interpretation B}}@
def digits(n):
    digits = [int(d) for d in str(n)]
    return math.prod(digits)
\end{lstlisting}
        \caption{Divergent implementations.}
        \label{fig:vagueness_ambiguity_implementations}
    \end{subfigure}
    \caption{A vagueness ambiguity example from Orchid-HEval task \#131, where the term ``digits'' is unspecified.}
    \label{fig:vagueness_overall_analysis}
\end{figure}

\begin{figure*}
  \centering
  \includegraphics[width=1\linewidth]{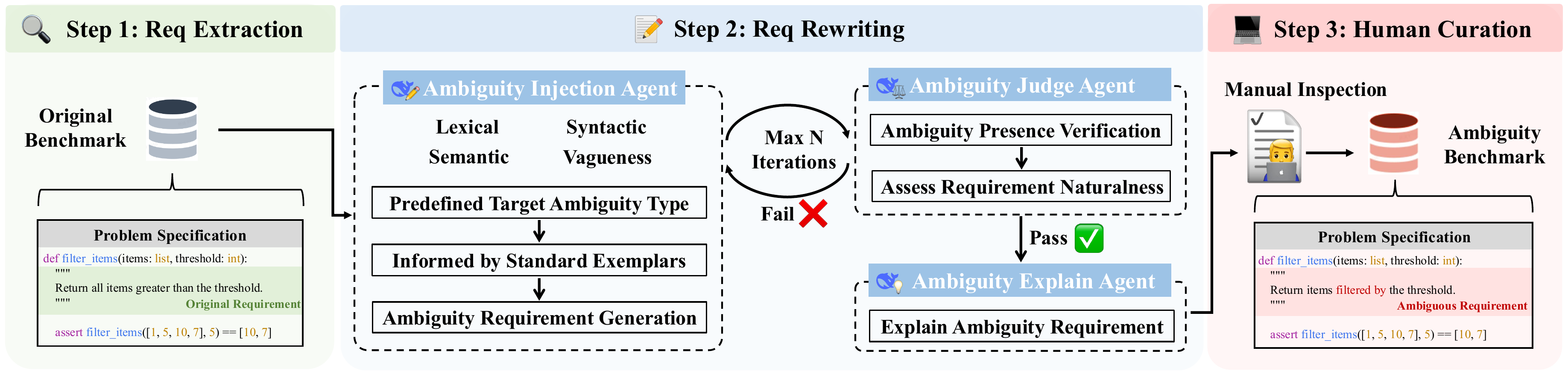}
  \caption{\textbf{Overview of \bench Construction Process.}}
  \label{fig:ambiguity_benchmark_construction}
\end{figure*}

Vagueness ambiguity arises when a requirement includes expressions that omit necessary details or lack sufficient specificity, leading to incomplete information. This deficiency permits multiple plausible interpretations within the requirement's context, resulting in varied understandings and implementations. Addressing vagueness ambiguity typically involves supplementing missing information or clarifying constraints.

Figure~\ref{fig:vagueness_ambiguity_example} illustrates an example of vagueness ambiguity. The requirement does not specify whether the product should be computed over all digits or only a subset, leaving the interpretation of which digits to include unclear. This lack of specificity permits two implementations: Implementation A (Figure~\ref{fig:vagueness_ambiguity_implementations}) multiplies only the odd digits, whereas Implementation B (Figure~\ref{fig:vagueness_ambiguity_implementations}) multiplies all digits regardless of parity. For test case \texttt{n = 3526}, implementation A returns \texttt{15}, while implementation B returns \texttt{180}.

%% file: sections/4-benchmark_design_and_dataset.tex
\begin{figure*}
    \centering
    \begin{minipage}[c]{0.6\linewidth} 
        \centering
        \resizebox{\linewidth}{!}{%
        \begin{threeparttable}
            \renewcommand{\arraystretch}{1.2} 
            \setlength{\tabcolsep}{5pt} 
            \begin{tabular}{l c !{\vrule width 0.4pt} c !{\vrule width 0.4pt} c c c c !{\vrule width 0.4pt} c c}
            \toprule
            \multicolumn{9}{c}{\bfseries Token Length} \\
            \rowcolor{gray!15}
            \textbf{Subset} & \textbf{\# (Ori / Amb)} & \textbf{Original} & \textbf{Lexical} & \textbf{Syntactic} & \textbf{Semantic} & \textbf{Vagueness} & \textbf{Avg Amb} & \textbf{Avg $\Delta$ (\%)} \\
            \midrule
            Orchid-HEval      & 164 / 656   & 143.47 & 140.23 & 140.12 & 140.78 & 135.87 & 139.25 & -2.94 \\
            Orchid-BCB         & 164 / 656   & 181.26 & 178.18 & 179.16 & 185.09 & 176.90 & 179.83 & -0.79 \\
            Orchid-BCB-Expand  & 976 / 3904  & 175.84 & 173.49 & 174.25 & 180.27 & 171.14 & 174.79 & -0.60 \\
            \midrule
            \multicolumn{9}{c}{\bfseries Perplexity} \\
            \rowcolor{gray!15}
            \textbf{Subset} & \textbf{\# (Ori / Amb)} & \textbf{Original} & \textbf{Lexical} & \textbf{Syntactic} & \textbf{Semantic} & \textbf{Vagueness} & \textbf{Avg Amb} & \textbf{Avg $\Delta$ (\%)} \\
            \midrule
            Orchid-HEval       & 164 / 656   & 30.55 & 37.39 & 37.04 & 32.81 & 36.72 & 35.99 & +17.78 \\
            Orchid-BCB         & 164 / 656   & 23.17 & 24.11 & 24.23 & 24.67 & 23.83 & 24.21 & +4.49 \\
            Orchid-BCB-Expand  & 976 / 3904  & 24.99 & 26.51 & 25.93 & 26.51 & 25.92 & 26.22 & +4.92 \\
            \bottomrule
            \end{tabular}
            \begin{tablenotes}
            \footnotesize
            \item \textbf{Note:} Avg $\Delta$ = (Avg Amb $-$ Original) / Original $\times 100\%$.
            \end{tablenotes}
        \end{threeparttable}%
        }
    \end{minipage}%
    \hfill
    \begin{minipage}[c]{0.4\linewidth}
        \centering
        \makebox[\linewidth][c]{
        \begin{tikzpicture}[font=\sffamily\small, scale=0.65, transform shape]
            \definecolor{coreBlue}{HTML}{4E79A7}
            \definecolor{coreSky}{HTML}{A0CBE8}
            \definecolor{coreGreen}{HTML}{59A14F}
            \definecolor{lineGray}{HTML}{999999}
            \def\Rmid{2.1}
            \def\Rin{1.25}
            \def\Rarm{0.4}
            \fill[coreBlue, draw=white, line width=1.5pt] (90:\Rmid) arc (90:-179.4:\Rmid) -- (-179.4:\Rin) arc (-179.4:90:\Rin) -- cycle;
            \fill[coreSky, draw=white, line width=1.5pt] (-179.4:\Rmid) arc (-179.4:-224.7:\Rmid) -- (-224.7:\Rin) arc (-224.7:-179.4:\Rin) -- cycle;
            \fill[coreGreen, draw=white, line width=1.5pt] (-224.7:\Rmid) arc (-224.7:-270:\Rmid) -- (-270:\Rin) arc (-270:-224.7:\Rin) -- cycle;
            \node[align=center, text=black!80] at (0,0) {\textbf{Orchid}\\[0.2em] \scriptsize Total Tasks\\[0.1em] \scriptsize 1,304};
            \draw[lineGray, thick] (-45:\Rmid) -- ++(-45:\Rarm) -- ++(0.3,0) coordinate (L1);
            \node[anchor=west, align=left, inner sep=3pt] at (L1) {\textbf{Orchid-BCB-Expand}\\ \textcolor{coreBlue}{\textbf{74.8\%}} \scriptsize (976)};
            \draw[lineGray, thick] (-200:\Rmid) -- ++(-200:\Rarm) -- ++(-0.3,0) coordinate (L2);
            \node[anchor=east, align=right, inner sep=3pt] at (L2) {\textbf{Orchid-HEval}\\ \textcolor{coreSky}{\textbf{12.6\%}} \scriptsize (164)};
            \draw[lineGray, thick] (-250:\Rmid) -- ++(-250:\Rarm) -- ++(-0.3,0) coordinate (L3);
            \node[anchor=east, align=right, inner sep=3pt] at (L3) {\textbf{Orchid-BCB}\\ \textcolor{coreGreen}{\textbf{12.6\%}} \scriptsize (164)};
            \node[align=center, font=\footnotesize, text=black] at (0, -3.0) {\textbf{Ambiguity Types}\\[0.3em] $\bullet$ Lexical \enspace $\bullet$ Syntactic \enspace $\bullet$ Semantic \enspace $\bullet$ Vagueness};
        \end{tikzpicture}
        }
    \end{minipage}
    \caption{\textbf{Orchid Benchmark Statistics.} The chart displays the distribution of data sources and lists the covered ambiguity types, while the table details token length and perplexity for original and ambiguous requirements.}
    \label{fig:orchid_statiscs}
\end{figure*}
\subsection{Methodology}
\label{Benchmark Construction}
\bench is constructed using the semi-automated human-in-the-loop pipeline shown in Figure~\ref{fig:ambiguity_benchmark_construction}. The pipeline comprises three stages: requirement extraction, requirement rewriting, and human curation. Before rewriting, we specify the target ambiguity type for each generation request. For each source requirement, we produce one \textit{lexical}, one \textit{syntactic}, one \textit{semantic}, and one \textit{vagueness} variant. We use the source programming task as the unit of construction. Each original requirement and its four variants form a matched set, allowing model behavior to be compared across ambiguity conditions while holding the source task constant.

\subsubsection{Requirement Extraction}

To prepare inputs for ambiguity rewriting, we first extract function-level requirements from existing datasets (green part in Figure~\ref{fig:ambiguity_benchmark_construction}). We use a rule-based parser to extract the natural language, and remove implementation-specific elements (\eg, input/output examples or code snippets), as clear requirements for rewriting.

\subsubsection{Requirement Rewriting}

We inject ambiguity into each source requirement using the multi-agent framework shown in Figure~\ref{fig:ambiguity_benchmark_construction}. The framework consists of an \textit{Ambiguity Injection Agent}, an \textit{Ambiguity Judge Agent}, and an \textit{Ambiguity Explanation Agent}, all implemented using DeepSeek-V3~\cite{liu2024deepseek}. The rewriting process aims to introduce a contextually plausible ambiguity while preserving the original functional intent.

First, the \textit{Ambiguity Injection Agent} rewrites each requirement according to a predefined target type. The target type is provided as input rather than inferred from the generated requirement. Guided by its definition and representative in-context examples~\cite{brown2020language}, the agent introduces one representative instance of the designated ambiguity while avoiding the other ambiguity types.

Next, the \textit{Ambiguity Judge Agent} evaluates each candidate according to three criteria: type specificity, implementation impact, and naturalness. Based on these assessments, the judge determines whether the candidate passes machine screening. If a candidate fails, the injection and evaluation steps are repeated for up to five iterations. If no candidate passes within this limit, the highest-rated candidate is forwarded to human curation rather than accepted automatically.

Finally, the \textit{Ambiguity Explanation Agent} identifies the ambiguous expression in the selected candidate and summarizes its plausible interpretations in a consistent format. The explanation is retained with the candidate to support subsequent human inspection and downstream analysis.

\subsubsection{Human Curation}
\label{sec:human-curation}

As illustrated in Figure~\ref{fig:ambiguity_benchmark_construction}, human curation constitutes the final stage of our construction pipeline for benchmark subsets designated for item-level validation. During this stage, the generated variants are partitioned among the curators, and each curator compares every assigned rewritten requirement with its corresponding original requirement.

Each variant is assessed against three criteria. First, the ambiguity introduced by the rewritten wording must be consistent with the intended ambiguity type. Second, the original intended behavior must remain a plausible interpretation of the rewritten requirement, and the underlying programming task must remain unchanged apart from the targeted ambiguity. Third, the rewritten requirement must support at least two plausible interpretations that could lead to functionally different implementations, rather than merely different wording or implementation styles.

Variants that fail any criterion are returned for requirement revision or ambiguity reinjection, as appropriate. Revised variants are inspected again using the same criteria and retained only after satisfying all three.

\subsection{\bench Instantiation}
\label{sec:benchmark-instantiation}

We instantiated the preceding construction procedure using two established function-level code-generation benchmarks: HumanEval+ \cite{liu2023your} and BigCodeBench~\cite{zhuo2024bigcodebench}. Orchid-HEval includes all 164 tasks from HumanEval+. To construct a size-matched core subset, we selected the first 164 tasks from BigCodeBench to form Orchid-BCB. For each task in these two subsets, the requirement-rewriting procedure generated four ambiguity variants.

Together, Orchid-HEval and Orchid-BCB constitute the manually curated core of \bench, comprising 328 source tasks and 1,312 ambiguity variants. All 1,312 variants underwent the item-level curation protocol described in Section~\ref{sec:human-curation}. The inspection and subsequent revision of these variants required over 246 person-hours.

To extend the scope of our evaluation beyond the 328-task manually curated core, we applied the same requirement-rewriting procedure to the remaining 976 BigCodeBench tasks. This produced 3,904 additional variants, collectively referred to as Orchid-BCB-Expand, enabling a larger-scale evaluation of model behavior under ambiguous requirements. Because this extension did not undergo the item-level manual curation applied to Orchid-HEval and Orchid-BCB, we report its results separately to make the difference in validation coverage explicit. Overall, \bench comprises 1,304 source tasks and 5,216 ambiguity variants across the three subsets.

\subsection{Benchmark Statistics}

As summarized in Figure~\ref{fig:orchid_statiscs}, \bench covers 1,304 tasks and 5,216 ambiguous requirements across lexical, syntactic, semantic, and vagueness. It comprises Orchid-HEval and Orchid-BCB, with 164 tasks each, and Orchid-BCB-Expand with 976.

Ambiguity causes only minor variations in token-level requirement length.
In Orchid-HEval, Orchid-BCB, and Orchid-BCB-Expand, the average token length decreases by 2.94\%, 0.79\%, and 0.60\%, respectively. These variations indicate that the overall structural characteristics of the tasks are preserved. Among ambiguity types, vagueness variants tend to be slightly shorter, whereas semantic variants are relatively longer. 

Ambiguity also increases linguistic uncertainty in requirements. In all subsets, ambiguous requirements consistently exhibit higher perplexity than originals, with an average increase of 17.78\% in Orchid-HEval, 4.49\% in Orchid-BCB, and 4.92\% in Orchid-BCB-Expand. Lexical and syntactic variants generally contribute more significantly to the perplexity increase, while semantic and vagueness variants lead to smaller, though still notable, increases.

\begin{table}
\centering
\caption{Selected LLMs.}
\label{tab:selected_llms}
\setlength{\tabcolsep}{2.5pt}
\renewcommand{\arraystretch}{1.2}
\resizebox{\linewidth}{!}{%
\begin{tabular}{llclcc}
\toprule
\textbf{Category} & \textbf{Models} & \textbf{Size} & \textbf{Publisher} & \textbf{Open Source} & \textbf{Release} \\
\midrule
\multirow{3}{*}{\textbf{General}} & GPT-4~\cite{achiam2023gpt} & N/A & OpenAI & No & Mar 2023 \\
 & DeepSeek-V3~\cite{liu2024deepseek} & 671B & DeepSeek & Yes & Dec 2024 \\
 & Claude-3.5~\cite{anthropic2024claude35} & N/A & Anthropic & No & Jun 2024 \\
\midrule
\multirow{2}{*}{\textbf{Code}} & CodeLlama-34B~\cite{roziere2023code} & 34B & Meta & Yes & Aug 2023 \\
 & Qwen-2.5-Coder~\cite{hui2024qwen2} & 32B & Alibaba & Yes & Sep 2024 \\
\midrule
\textbf{Reasoning} & DeepSeek-R1~\cite{guo2025deepseek} & 671B & DeepSeek & Yes & Jan 2025 \\
\bottomrule
\end{tabular}%
}
\end{table}

%% file: sections/5-the_impact_of_requirement_ambiguity.tex
\begin{table*}
\caption{Pass@$k$ (\%) on original and paired changes. Each original requirement and its four ambiguity variants are evaluated as a paired group using the same model configuration and test suite.}
\label{tab:combined-ambiguity-results-expand}
\centering
\resizebox{0.98\textwidth}{!}{%
\begin{threeparttable}
\begin{tabular}{l!{\vrule}c!{\vrule}r r r r!{\vrule}c!{\vrule}r r r r!{\vrule}c!{\vrule}r r r r}

\hline
\multirow{2}{*}{\textbf{Models}} 
& \multicolumn{5}{c!{\vrule}}{\textbf{Orchid-HEval}} 
& \multicolumn{5}{c!{\vrule}}{\textbf{Orchid-BCB}} 
& \multicolumn{5}{c}{\textbf{Orchid-BCB-Expand}} \\
\cmidrule(lr){2-6} \cmidrule(lr){7-11} \cmidrule(lr){12-16}
& \textbf{Orig.} & $\Delta$ Lex & $\Delta$ Syn & $\Delta$ Sem & $\Delta$ Vag
& \textbf{Orig.} & $\Delta$ Lex & $\Delta$ Syn & $\Delta$ Sem & $\Delta$ Vag
& \textbf{Orig.} & $\Delta$ Lex & $\Delta$ Syn & $\Delta$ Sem & $\Delta$ Vag\\
\hline
\rowcolor[rgb]{0.9,0.9,0.9}
\multicolumn{16}{c}{\textbf{Pass@1 (\%)}} \\
\hline
CodeLlama & \textbf{33.66} & -4.39 & -2.93 & -2.32 & -7.07 & \textbf{6.22} & -1.34 & -1.34 & -1.59 & -0.49 & \textbf{5.72} & -0.84 & -0.93 & -1.42 & -0.58 \\
Qwen-2.5-Coder & \textbf{69.15} & -6.59 & -3.78 & -2.93 & -9.15 & \textbf{42.80} & -3.17 & 1.71 & -7.56 & 0.13 & \textbf{44.67} & -1.41 & -1.12 & -6.49 & -3.38 \\
DeepSeek-V3 & \textbf{81.22} & -4.02 & -2.32 & -7.07 & -11.46 & \textbf{48.05} & -8.78 & -0.37 & -5.98 & -3.98 & \textbf{50.04} & -3.69 & -2.07 & -7.46 & -5.00 \\
Claude-3.5 & \textbf{76.71} & -8.66 & -31.10 & -2.44 & -8.78 & \textbf{41.22} & -6.71 & 0.24 & -4.04 & -3.44 & \textbf{45.16} & -0.84 & -1.70 & -6.50 & -3.08 \\
GPT-4 & \textbf{72.68} & -8.78 & -3.66 & -6.34 & -12.56 & \textbf{45.24} & -29.14 & -28.65 & -30.12 & -29.26 & \textbf{47.62} & -28.38 & -26.94 & -32.35 & -28.67 \\
DeepSeek-R1 & \textbf{77.68} & -1.09 & -1.58 & -4.02 & -4.39 & \textbf{33.90} & -4.27 & 0.04 & -10.12 & -2.31 & \textbf{31.42} & -2.14 & -2.08 & -9.70 & -2.84\\
\hline
\rowcolor[rgb]{0.9,0.9,0.9}
\multicolumn{16}{c}{\textbf{Pass@3 (\%)}} \\
\hline
CodeLlama & \textbf{40.49} & -4.64 & -2.93 & -1.16 & -8.36 & \textbf{15.73} & -3.17 & -2.32 & -3.53 & -1.46 & \textbf{14.40} & -2.30 & -1.77 & -3.30 & -1.15 \\
Qwen-2.5-Coder & \textbf{70.55} & -6.40 & -2.44 & -1.53 & -7.14 & \textbf{47.56} & -2.13 & 2.87 & -4.76 & 0.51 & \textbf{50.91} & -0.61 & -0.66 & -4.02 & -2.46 \\
DeepSeek-V3 & \textbf{83.41} & -2.50 & -2.62 & -7.43 & -11.40 & \textbf{53.48} & -9.33 & -0.31 & -3.00 & -2.41 & \textbf{58.26} & -3.01 & -1.20 & -6.31 & -3.80 \\
Claude-3.5 & \textbf{89.94} & -8.23 & -26.34 & -6.83 & -9.45 & \textbf{46.28} & -5.91 & 1.77 & -4.94 & -2.62 & \textbf{49.48} & -0.68 & -1.17 & -4.97 & -2.53 \\
GPT-4 & \textbf{82.68} & -10.85 & -4.14 & -8.11 & -14.39 & \textbf{54.02} & -29.81 & -29.08 & -30.97 & -31.46 & \textbf{56.71} & -28.11 & -26.37 & -33.44 & -28.95 \\
DeepSeek-R1 & \textbf{86.83} & -1.46 & -1.59 & -3.66 & -4.21 & \textbf{44.08} & -6.34 & 0.01 & -11.11 & -2.74 & \textbf{43.10} & -2.66 & -3.09 & -10.97 & -4.03 \\
\hline
\rowcolor[rgb]{0.9,0.9,0.9}
\multicolumn{16}{c}{\textbf{Pass@5 (\%)}} \\
\hline
CodeLlama & \textbf{43.29} & -3.66 & -2.44 & -0.61 & -8.53 & \textbf{23.17} & -4.88 & -1.83 & -4.88 & -2.44 & \textbf{20.70} & -3.20 & -2.65 & -4.51 & -1.17 \\
Qwen-2.5-Coder & \textbf{70.73} & -6.10 & -1.83 & -1.22 & -6.10 & \textbf{48.78} & -1.83 & 3.66 & -3.05 & -0.36 & \textbf{52.56} & -0.10 & -0.41 & -2.56 & -2.05 \\
DeepSeek-V3 & \textbf{84.76} & -2.44 & -3.66 & -8.54 & -12.20 & \textbf{54.88} & -9.15 & -0.40 & -1.83 & -1.51 & \textbf{61.17} & -2.97 & -1.03 & -5.84 & -3.69 \\
Claude-3.5 & \textbf{92.07} & -4.87 & -22.56 & -6.70 & -8.53 & \textbf{48.17} & -6.10 & 1.83 & -4.83 & -2.82 & \textbf{50.72} & -0.42 & -0.62 & -2.84 & -1.69 \\
GPT-4 & \textbf{86.59} & -12.81 & -4.88 & -9.76 & -16.47 & \textbf{57.31} & -29.87 & -28.65 & -30.48 & -31.09 & \textbf{60.04} & -27.36 & -25.72 & -33.20 & -28.34 \\
DeepSeek-R1 & \textbf{90.24} & -2.44 & -2.44 & -3.04 & -6.09 & \textbf{48.78} & -6.71 & -1.05 & -11.74 & -3.66 & \textbf{47.38} & -2.20 & -3.92 & -10.90 & -4.32 \\
\hline
\end{tabular}
\begin{tablenotes}
\item * The abbreviation Orig. stands for original requirement (\ie, without ambiguity).
\item* The abbreviations Lex, Syn, Sem and Vag stand for lexical, syntactic, semantic and vagueness ambiguity, respectively.
\item * $\Delta=\mathrm{Pass@}k_{\mathrm{ambiguous}}-\mathrm{Pass@}k_{\mathrm{original}}$.

\end{tablenotes}
\end{threeparttable}%
}
\end{table*}

\subsection{Experimental Setup}
\noindent\textbf{LLM Selection.}
We adopt a series-representative strategy to efficiently cover major model families while balancing capability and computational cost. As shown in Table~\ref{tab:selected_llms}, we select six LLMs, covering three categories of general-purpose, code-specialized, and reasoning models
across diverse parameter sizes.

\noindent\textbf{Evaluation Protocol.}
We use the source benchmark task as the unit of analysis. For each task, we evaluate the original requirement and its four corresponding variants, one for each ambiguity type. For each combination of model and benchmark, the original requirement and its variants use the same prompt wrapper, generation configuration, and source-task tests. We generate $n=5$ completions for every model and requirement pair. Therefore, each ambiguity variant is compared with its corresponding original requirement over the same set of source tasks, rather than with an unrelated set of clear requirements. For each ambiguity
type, we define $\Delta=\mathrm{Pass@}k_{\mathrm{ambiguous}}-\mathrm{Pass@}k_{\mathrm{original}}$, where a negative value indicates performance degradation.

\noindent\textbf{LLM Settings.}
We retain the benchmark-specific prompt formats to preserve comparability with the original evaluations. For each model and benchmark, the original requirement and its four ambiguity variants are evaluated under identical generation settings, ensuring a controlled comparison of requirement wording.

\noindent\textbf{Test Suites.}
For Orchid-HEval, we use an EvalPlus-based evaluation harness with the HumanEval+ tests~\cite{liu2023your}. For Orchid-BCB and Orchid-BCB-Expand, we use the tests associated with the corresponding BigCodeBench source tasks~\cite{zhuo2024bigcodebench}. The same tests and execution limits are applied to each original requirement and its corresponding variants. Syntax errors, runtime exceptions, and timeouts are counted as failures.

\noindent\textbf{Evaluation Metrics.}
We use the following metrics:
\begin{itemize}[leftmargin=*]
    \item \textit{Pass@$k$}: The estimated probability that at least one of the $k$ generated code samples passes all unit tests: $\mathrm{Pass@}k = 1-\binom{n-c}{k}/\binom{n}{k}$, where $c$ is the number of correct samples among $n=5$ generations. We report the average across tasks for $k\in\{1,3,5\}$.

    \item \textit{Conflict Rate}: The proportion of response pairs whose test-outcome vectors differ on at least one test: $\mathrm{ConflictRate}=C/\binom{r}{2}$, where $C$ is the number of conflicting pairs among the $r$ responses. We compute the conflict rate for each task and average across tasks.

    \item \textit{Ambiguity Recognition}: We assess whether a model can recognize, localize, and address requirement ambiguity using four ascending levels: (i) \textit{Unaware}, where the model does not recognize ambiguity; (ii) \textit{Detection}, where it recognizes ambiguity but does not identify its source; (iii) \textit{Localization}, where it correctly identifies the ambiguous segment; and (iv) \textit{Tackling}, where it provides plausible interpretations or clarification options. Detection, Localization, and Tackling are treated as positive ambiguity detections when computing precision and recall.
\end{itemize}

\subsection{RQ1: Performance Impact}
Table~\ref{tab:combined-ambiguity-results-expand} summarizes the Pass@$k$ results. Overall, ambiguity poses a pervasive challenge to reliable code generation. It consistently degrades generation quality across all evaluated models, reducing Pass@1 accuracy by an average of 7.22 percentage points, with the largest observed decline reaching 32.35 points.

Notably, strong performance on clear requirements does not necessarily translate into stable behavior under ambiguous inputs. \bench is effective in revealing such latent capability gaps that are not captured by standard benchmarks. For example, although GPT-4 achieves top-tier baseline performance, its accuracy drops by more than 28 percentage points on Orchid-BCB under ambiguous requirements. In contrast, open-source models such as Qwen-2.5-Coder exhibit relatively higher stability, with performance degradation limited to approximately 8 percentage points.

In addition, \bench reveals fine-grained intra-model sensitivity to different types of ambiguity. For instance, Claude-3.5 is highly affected by syntactic ambiguity, with a performance drop of 31.10 percentage points, while its performance remains relatively stable under semantic ambiguity, with a decline of only 2.44 percentage points. These results indicate that ambiguity affects models in a non-uniform manner and interacts with both model-specific characteristics and ambiguity types. Overall, these findings highlight the necessity of \bench for a comprehensive evaluation of LLMs under realistic and ambiguous requirements.

\begin{myboxc} 
\textbf{Finding 1:} High performance on clear requirements does not guarantee stability under ambiguity. On average, models suffer a relative performance decline of 16.25\% in Pass@1 accuracy.
\end{myboxc}

\subsection{RQ2: Functional Consistency}

We evaluate functional consistency by calculating the average conflict rate across five responses per model (intra-model) and across all responses from five models (inter-model).

As summarized in Table~\ref{tab:ambig-analysis}, ambiguous requirements substantially increase functional divergence in LLM-generated code. While clear requirements maintain relatively higher consistency, ambiguity nearly doubles the conflict rates for capable models on Orchid-HEval. Specifically, GPT-4 increases from 14.09\% to 28.29\%, and DeepSeek-V3 from 6.83\% to 17.45\%. This trend persists across ambiguity types; for example, lexical ambiguity alone increases Qwen-2.5-Coder's conflict rate on Orchid-BCB from 16.89\% to 21.52\%. 

Notably, CodeLlama does not exhibit a similar increase, as its Pass@1 of only 6.22\% limits the availability of correct outputs, making consistency comparison infeasible. Furthermore, ambiguity induces substantial divergence across models. On Orchid-BCB, the inter-model conflict rate reaches 57.28\%, indicating a lack of consensus among LLMs when interpreting ambiguous requirements.

\begin{figure*}
  \centering
  \includegraphics[width=0.87\linewidth]{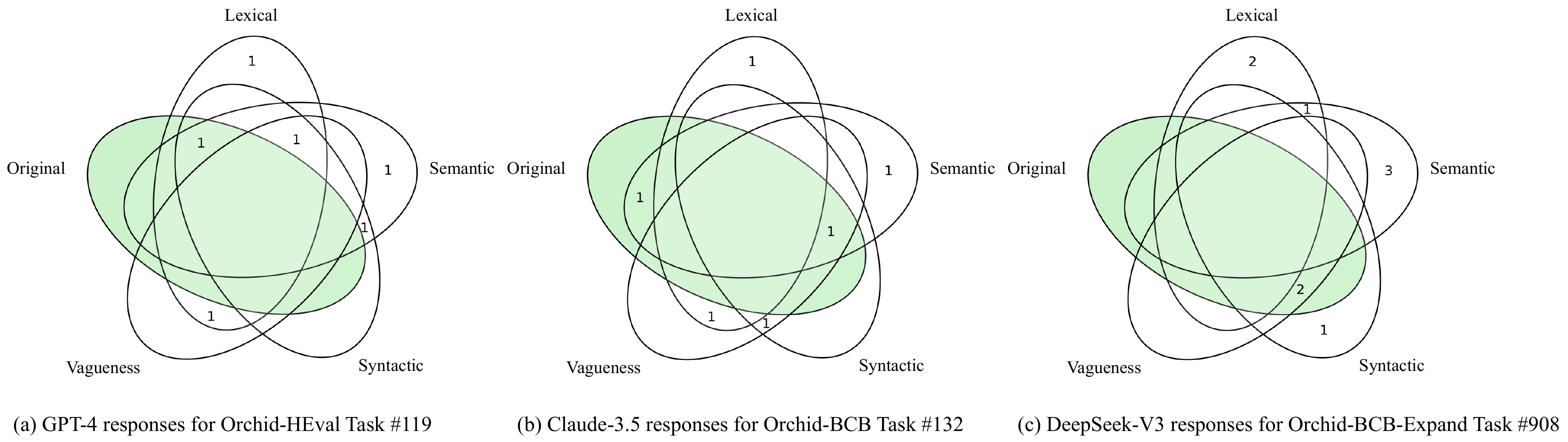}
    \caption{Functional diversity of LLMs on \bench.}
  \label{fig:venn_figue}
\end{figure*}

Figure~\ref{fig:venn_figue} further illustrates the functional fragmentation introduced by ambiguity. The central green region represents the functionality derived from clear requirements, while the separated regions indicate that ambiguity leads to multiple incompatible functional interpretations. For instance, in Orchid-HEval Task \#119 (Figure~\ref{fig:venn_figue}a), GPT-4 generates up to five distinct functional variants, resulting in implementations that diverge significantly from the intended behavior. Similar patterns are observed across other models, as shown in Figures~\ref{fig:venn_figue}b and \ref{fig:venn_figue}c, confirming that ambiguity consistently undermines functional consistency.

\begin{myboxc} \textbf{Finding 2:} Ambiguous requirements lead LLMs to generate functionally inconsistent code. For instance, ambiguity nearly doubles GPT-4's intra-model conflict rate to 28.29\% and increases the inter-model conflict rate from 51.04\% to 57.28\%.
\end{myboxc}
\begin{table*}
\centering
\caption{Conflict Rate of LLMs on original and ambiguous requirements.}
\label{tab:ambig-analysis}
\small
\setlength{\tabcolsep}{5pt}
\setlength{\heavyrulewidth}{0.06em}
\setlength{\lightrulewidth}{0.03em}
\setlength{\abovetopsep}{0pt}
\begin{threeparttable}
\begin{tabular*}{0.95\textwidth}{@{\extracolsep{\fill}}l cc c cc c cc}
\toprule
\multirow{2}{*}{\textbf{Models}} & \multicolumn{2}{c}{\textbf{Orchid-HEval (\%)}} && \multicolumn{2}{c}{\textbf{Orchid-BCB (\%)}} && \multicolumn{2}{c}{\textbf{Orchid-BCB-Expand (\%)}} \\
\cmidrule(lr){2-3} \cmidrule(lr){5-6} \cmidrule(lr){8-9}
& Original & Ambiguous && Original & Ambiguous && Original & Ambiguous \\
\midrule
CodeLlama      & 23.96 & \textbf{38.61} && \textbf{50.12} & 47.62 && \textbf{48.24} & 45.93 \\
Qwen-2.5-Coder & 5.67  & \textbf{22.07} && 16.89 & \textbf{28.06} && 18.49 & \textbf{31.16} \\
DeepSeek-V3    & 6.83  & \textbf{17.45} && 17.07 & \textbf{26.23} && 21.73 & \textbf{31.62} \\
Claude-3.5     & 16.83 & \textbf{21.16} && 12.68 & \textbf{24.01} && 13.10 & \textbf{22.48} \\
GPT-4          & 14.09 & \textbf{28.29} && 22.80 & \textbf{31.42} && 24.70 & \textbf{35.33} \\
\midrule
Multi-models   & 30.60 & \textbf{36.45} && 51.04 & \textbf{57.28} && 54.30 & \textbf{58.56} \\
\bottomrule
\end{tabular*}
\begin{tablenotes}
\item * \textbf{Bold} indicates the higher conflict rate between Original and Ambiguous.
\end{tablenotes}
\end{threeparttable}
\end{table*}

\subsection{RQ3: Identification and Resolution}
We evaluate each model on both the original clear requirements and their ambiguous variants using a fixed prompt that asks whether ambiguity is present and, if so, requests an explanation without generating code. GPT-4 (\texttt{gpt-4}) then compares each response with the corresponding reference annotation and assigns one of the four recognition levels defined above. The judge receives only the model response and reference annotation, without the identity of the evaluated model. The evaluated API models and the judge use a temperature of 0.0, while local models use greedy decoding. \textit{Detection}, \textit{Localization}, and \textit{Tackling} are counted as positive ambiguity detections when computing precision and recall. We manually inspect a random subset of 50 responses and observe 96\% agreement between the automatic and human judgments.

\begin{figure*}[h]
  \centering
  \includegraphics[width=0.97\linewidth]{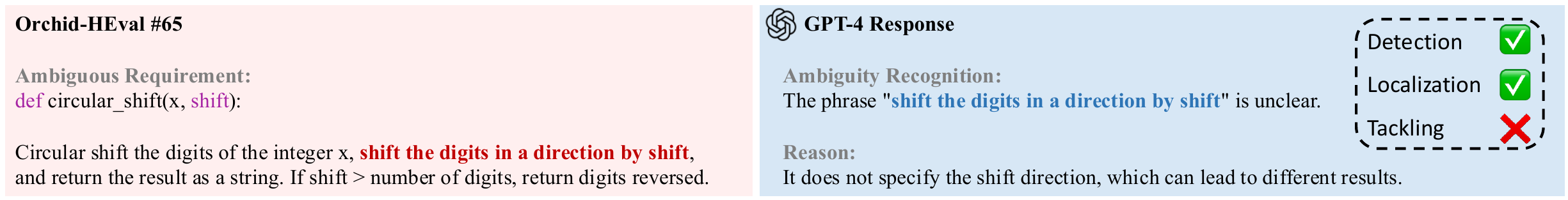}
  \caption{Example from Orchid-HEval \#65 where GPT-4 recognizes and localizes ambiguity.}
  \label{fig:ambiguity_recognition}
  \vspace{-0.2cm}
\end{figure*}

Figure~\ref{fig:ambiguity_recognition} shows a representative example from Orchid-HEval \#65, which involves circularly shifting the digits of an integer $x$. The phrase ``shift the digits in a direction by shift'' is inherently ambiguous due to the unspecified direction, leading to multiple possible interpretations and outputs. In this case, GPT-4 successfully detects and localizes the ambiguous segment but does not provide a concrete resolution. Accordingly, its response is categorized as successful localization only.

Table~\ref{tab:ambiguity-breakdown-combined-clean-expand-precise} summarizes the evaluation results. While recall varies across models, precision remains consistently around 50\%, indicating that clear requirements are frequently misclassified as ambiguous. Overall, all evaluated LLMs struggle to reliably distinguish ambiguity from complex but well-defined requirements, and tend to adopt a conservative strategy that favors over-detection. This results in a high false positive rate and consequently low precision.

Among the evaluated models, Claude-3.5 exhibits the most pronounced behavior. By adopting a highly sensitive detection strategy to minimize missed ambiguities, it achieves near-perfect recall (often exceeding 96\%, and reaching 100\% for semantic ambiguity). However, this comes at the expense of precision, which remains around 50\%, indicating that a substantial portion of clear requirements are incorrectly classified as ambiguous.

\begin{myboxc} \textbf{Finding 3:} LLMs struggle to distinguish ambiguous requirements. They tend to regard most requirements as ambiguous (precision around 50\%).
\end{myboxc}

We further analyze the levels of LLMs' ambiguity recognition in Table~\ref{tab:ambiguity-breakdown-combined-clean-expand-precise}. For each case where a model identifies ambiguity, its response is categorized into three progressive levels: detection (Det), localization (Loc), and tackling (Tac). Overall, while LLMs demonstrate reasonable capability in detecting ambiguous requirements, their ability to localize and tackle ambiguity remains limited.

On Orchid-BCB, this gap is evident across models. For example, Claude-3.5 detects 73.2\% of syntactic ambiguities, yet achieves only 7.9\% in localization and 17.1\% in tackling. Similarly, GPT-4 detects 43.9\% of vagueness cases, but attains a localization rate of 15.2\% and a tackling rate of 21.3\%. These results indicate that although ambiguity can often be identified, transforming detection into precise localization and actionable resolution remains challenging.

We further observe that recognition performance varies across ambiguity types. Lexical ambiguities are detected at 40.9\%, syntactic ambiguities at 43.7\%, semantic ambiguities at 24.7\%, and vagueness at 38.8\%, while localization rates are 11.1\%, 13.7\%, 12.5\%, and 11.1\%, respectively. In contrast, semantic ambiguities achieve the highest conditional tackling rate across model-dataset combinations, with 47.0\% of recognized cases reaching the tackling level on average, suggesting that they are more amenable to resolution once identified than other ambiguity types.

\begin{myboxc} \textbf{Finding 4:} LLMs struggle to localize and tackle ambiguous requirements. Across evaluated models, localization is consistently poor, rarely exceeding 23\%, while tackling remains below 64\%.
\end{myboxc}

\newcommand{\blueheat}[1]{%
    \ifdim#1pt>90pt \cellcolor{skyblue!100}\hspace{2pt}#1\hspace{2pt}%
    \else\ifdim#1pt>80pt \cellcolor{skyblue!90}\hspace{2pt}#1\hspace{2pt}%
    \else\ifdim#1pt>70pt \cellcolor{skyblue!80}\hspace{2pt}#1\hspace{2pt}%
    \else\ifdim#1pt>60pt \cellcolor{skyblue!70}\hspace{2pt}#1\hspace{2pt}%
    \else\ifdim#1pt>50pt \cellcolor{skyblue!60}\hspace{2pt}#1\hspace{2pt}%
    \else\ifdim#1pt>40pt \cellcolor{skyblue!50}\hspace{2pt}#1\hspace{2pt}%
    \else\ifdim#1pt>30pt \cellcolor{skyblue!40}\hspace{2pt}#1\hspace{2pt}%
    \else\ifdim#1pt>20pt \cellcolor{skyblue!30}\hspace{2pt}#1\hspace{2pt}%
    \else\ifdim#1pt>10pt \cellcolor{skyblue!20}\hspace{2pt}#1\hspace{2pt}%
    \else \cellcolor{skyblue!10}\hspace{2pt}#1\hspace{2pt}%
    \fi\fi\fi\fi\fi\fi\fi\fi\fi
}
\begin{table*}
\caption{Evaluation of LLM capability of recognizing ambiguity in requirements.}

\label{tab:ambiguity-breakdown-combined-clean-expand-precise}
\renewcommand{\arraystretch}{0.95}
\setlength{\tabcolsep}{1pt}
\centering
\resizebox{0.96\textwidth}{!}{%
\begin{threeparttable}
\begin{tabular}{l!{\vrule width 0.4pt}c c!{\vrule width 0.4pt}c c c c!{\vrule width 0.4pt}c c!{\vrule width 0.4pt}c c c c!{\vrule width 0.4pt}c c!{\vrule width 0.4pt}c c c c}
\hline
\multirow{2}{*}{\textbf{Models}}
& \multicolumn{6}{c!{\vrule width 0.4pt}}{\textbf{Orchid-HEval}}
& \multicolumn{6}{c!{\vrule width 0.4pt}}{\textbf{Orchid-BCB}}
& \multicolumn{6}{c}{\textbf{Orchid-BCB-Expand}} \\
\cmidrule(lr){2-7} \cmidrule(lr){8-13} \cmidrule(lr){14-19}
& Pre(\%) & Rec(\%) & Una(\%) & Det(\%) & Loc(\%) & Tac(\%)
& Pre(\%) & Rec(\%) & Una(\%) & Det(\%) & Loc(\%) & Tac(\%)
& Pre(\%) & Rec(\%) & Una(\%) & Det(\%) & Loc(\%) & Tac(\%) \\
\hline
\rowcolor{gray!15}\multicolumn{19}{c}{\textbf{Lexical Ambiguity}} \\
\hline
GPT-4 & 55.2 & 78.7 & 21.3 & \blueheat{20.7} & \blueheat{13.5} & \blueheat{44.5}
& 47.4 & 79.3 & 20.8 & \blueheat{45.1} & \blueheat{14.6} & \blueheat{19.5}
& 48.3 & 84.0 & 16.0 & \blueheat{48.8} & \blueheat{15.5} & \blueheat{19.7} \\
Claude-3.5 & 50.8 & 97.0 & 3.0 & \blueheat{31.1} & \blueheat{6.1} & \blueheat{59.8}
& 49.7 & 97.6 & 2.4 & \blueheat{70.1} & \blueheat{5.5} & \blueheat{22.0}
& 50.5 & 99.7 & 0.3 & \blueheat{68.4} & \blueheat{9.7} & \blueheat{21.6} \\
DeepSeek-V3 & 55.4 & 78.0 & 22.0 & \blueheat{21.3} & \blueheat{11.0} & \blueheat{45.7}
& 50.5 & 89.0 & 11.0 & \blueheat{54.8} & \blueheat{5.5} & \blueheat{28.7}
& 50.1 & 88.7 & 11.3 & \blueheat{53.2} & \blueheat{11.2} & \blueheat{24.3} \\
Qwen-2.5-Coder & 51.8 & 87.8 & 12.2 & \blueheat{28.6} & \blueheat{17.7} & \blueheat{41.5}
& 55.9 & 89.6 & 10.4 & \blueheat{56.7} & \blueheat{12.8} & \blueheat{20.1}
& 51.8 & 87.8 & 12.2 & \blueheat{52.1} & \blueheat{16.5} & \blueheat{19.2} \\
CodeLlama & 57.1 & 39.0 & 61.0 & \blueheat{18.9} & \blueheat{12.8} & \blueheat{7.3}
& 42.0 & 35.4 & 64.6 & \blueheat{23.8} & \blueheat{6.7} & \blueheat{4.9}
& 39.2 & 30.6 & 69.4 & \blueheat{20.1} & \blueheat{7.2} & \blueheat{3.3} \\
\hline
\rowcolor{gray!15}\multicolumn{19}{c}{\textbf{Syntactic Ambiguity}} \\
\hline
GPT-4 & 51.8 & 68.9 & 31.1 & \blueheat{33.5} & \blueheat{15.3} & \blueheat{20.1}
& 48.7 & 83.5 & 16.5 & \blueheat{50.5} & \blueheat{16.5} & \blueheat{16.5}
& 48.9 & 85.8 & 14.2 & \blueheat{45.3} & \blueheat{21.7} & \blueheat{18.8} \\
Claude-3.5 & 50.0 & 93.9 & 6.1 & \blueheat{48.8} & \blueheat{12.2} & \blueheat{32.9}
& 49.8 & 98.1 & 1.8 & \blueheat{73.2} & \blueheat{7.9} & \blueheat{17.1}
& 49.6 & 98.9 & 1.1 & \blueheat{62.9} & \blueheat{14.6} & \blueheat{21.4} \\
DeepSeek-V3 & 53.2 & 71.3 & 28.7 & \blueheat{29.8} & \blueheat{12.2} & \blueheat{29.3}
& 50.7 & 89.7 & 10.4 & \blueheat{51.8} & \blueheat{16.5} & \blueheat{21.3}
& 50.5 & 89.6 & 10.5 & \blueheat{45.4} & \blueheat{16.8} & \blueheat{27.3} \\
Qwen-2.5-Coder & 49.6 & 80.5 & 19.5 & \blueheat{47.6} & \blueheat{15.2} & \blueheat{17.7}
& 54.7 & 85.4 & 14.6 & \blueheat{55.5} & \blueheat{15.9} & \blueheat{14.0}
& 51.6 & 86.6 & 13.4 & \blueheat{49.6} & \blueheat{20.9} & \blueheat{16.1} \\
CodeLlama & 49.5 & 28.7 & 71.3 & \blueheat{21.4} & \blueheat{4.3} & \blueheat{3.0}
& 38.0 & 29.9 & 70.1 & \blueheat{22.0} & \blueheat{6.1} & \blueheat{1.8}
& 39.9 & 31.6 & 68.4 & \blueheat{18.6} & \blueheat{9.2} & \blueheat{3.8} \\
\hline
\rowcolor{gray!15}\multicolumn{19}{c}{\textbf{Semantic Ambiguity}} \\
\hline
GPT-4 & 51.4 & 67.7 & 32.3 & \blueheat{15.9} & \blueheat{21.3} & \blueheat{30.5}
& 50.0 & 87.8 & 12.2 & \blueheat{24.4} & \blueheat{12.8} & \blueheat{50.6}
& 50.0 & 89.8 & 10.2 & \blueheat{25.1} & \blueheat{16.9} & \blueheat{47.8} \\
Claude-3.5 & 50.8 & 96.9 & 3.0 & \blueheat{29.9} & \blueheat{17.1} & \blueheat{50.0}
& 50.3 & 100.0 & 0.0 & \blueheat{32.9} & \blueheat{6.1} & \blueheat{61.0}
& 50.1 & 99.7 & 0.3 & \blueheat{36.8} & \blueheat{7.8} & \blueheat{55.1} \\
DeepSeek-V3 & 50.9 & 65.2 & 34.8 & \blueheat{15.2} & \blueheat{11.0} & \blueheat{39.0}
& 51.9 & 93.9 & 6.1 & \blueheat{24.4} & \blueheat{6.7} & \blueheat{62.8}
& 51.7 & 94.6 & 5.4 & \blueheat{22.2} & \blueheat{9.3} & \blueheat{63.1} \\
Qwen-2.5-Coder & 51.1 & 85.4 & 14.6 & \blueheat{32.9} & \blueheat{14.7} & \blueheat{37.8}
& 56.6 & 92.1 & 7.9 & \blueheat{26.3} & \blueheat{20.1} & \blueheat{45.7}
& 52.9 & 91.3 & 8.7 & \blueheat{28.3} & \blueheat{23.0} & \blueheat{40.0} \\
CodeLlama & 50.5 & 29.9 & 70.1 & \blueheat{20.1} & \blueheat{4.3} & \blueheat{5.5}
& 40.3 & 32.9 & 67.1 & \blueheat{21.3} & \blueheat{7.3} & \blueheat{4.3}
& 39.3 & 30.8 & 69.2 & \blueheat{15.5} & \blueheat{9.7} & \blueheat{5.6} \\
\hline
\rowcolor{gray!15}\multicolumn{19}{c}{\textbf{Vagueness Ambiguity}} \\
\hline
GPT-4 & 57.1 & 85.4 & 14.6 & \blueheat{28.6} & \blueheat{15.9} & \blueheat{40.9}
& 47.8 & 80.5 & 19.6 & \blueheat{43.9} & \blueheat{15.2} & \blueheat{21.3}
& 48.7 & 85.4 & 14.7 & \blueheat{42.8} & \blueheat{17.8} & \blueheat{24.7} \\
Claude-3.5 & 51.3 & 98.8 & 1.2 & \blueheat{28.7} & \blueheat{7.9} & \blueheat{62.2}
& 50.3 & 100.0 & 0.0 & \blueheat{65.2} & \blueheat{4.3} & \blueheat{30.5}
& 50.0 & 99.3 & 0.7 & \blueheat{56.5} & \blueheat{6.8} & \blueheat{36.0} \\
DeepSeek-V3 & 58.5 & 88.4 & 11.6 & \blueheat{26.2} & \blueheat{8.5} & \blueheat{53.7}
& 50.3 & 88.4 & 11.6 & \blueheat{51.8} & \blueheat{4.9} & \blueheat{31.7}
& 50.8 & 90.6 & 9.4 & \blueheat{42.3} & \blueheat{11.3} & \blueheat{37.0} \\
Qwen-2.5-Coder & 53.5 & 93.9 & 6.1 & \blueheat{32.3} & \blueheat{17.1} & \blueheat{44.5}
& 55.2 & 87.2 & 12.8 & \blueheat{52.4} & \blueheat{17.1} & \blueheat{17.7}
& 51.9 & 87.6 & 12.4 & \blueheat{48.8} & \blueheat{17.3} & \blueheat{21.5} \\
CodeLlama & 49.5 & 28.7 & 71.3 & \blueheat{16.5} & \blueheat{6.1} & \blueheat{6.1}
& 42.5 & 36.0 & 64.0 & \blueheat{23.8} & \blueheat{7.9} & \blueheat{4.3}
& 41.5 & 33.7 & 66.3 & \blueheat{22.7} & \blueheat{8.0} & \blueheat{3.0} \\
\hline
\end{tabular}
\begin{tablenotes}[flushleft]
\item * Darker blue cells indicate a higher percentage of responses for that category.
\end{tablenotes}
\end{threeparttable}
}
\end{table*}

%% file: sections/6-learned_lessons.tex
Based on our findings, we summarize several key lessons on how to handle ambiguous requirements, with particular emphasis on stability, consistency, and ambiguity recognition.

\paragraph{1) High Performance Does Not Necessarily Translate to Stability Under Ambiguity.}
Even state-of-the-art models, such as GPT-4, exhibit notable performance degradation when exposed to ambiguous requirements, whereas some models with comparatively lower baseline performance, such as Qwen-2.5-Coder, demonstrate relatively stable behavior across such inputs. This suggests that leaderboard performance on well-formed benchmarks is insufficient to characterize model effectiveness in realistic software engineering scenarios, where requirements are often underspecified or ambiguous. For practitioners, this implies that model selection should incorporate targeted evaluation under project-specific ambiguity patterns rather than relying solely on aggregate benchmark rankings.

\paragraph{2) Ambiguity Harms Both Intra- and Inter-Model Consistency.}
Requirement ambiguity increases variability not only across different models but also across multiple outputs from the same model. Such inconsistencies reflect uncertainty in interpreting the requirements and can serve as an empirical signal of underlying ambiguity. In practice, developers can leverage this property by generating multiple candidate solutions or comparing outputs across models. Significant divergence in functional or logical behavior should be treated as a warning sign, prompting further clarification of the requirements before proceeding with implementation.

\paragraph{3) Sensitivity to Ambiguity Is Type-Dependent.}
LLMs exhibit uneven sensitivity to different categories of ambiguity. Some models are more affected by syntactic or vague expressions, while others show relatively minor performance variation across ambiguity types. This indicates that ambiguity should not be treated as a uniform phenomenon when evaluating code generation systems. Instead, fine-grained analysis is necessary to understand model behavior under different ambiguity conditions. These type-specific differences also have practical implications for requirements engineering. For development teams, identifying recurring ambiguity patterns in internal requirements (e.g., syntactic ambiguity or vagueness) can inform the design of guidelines and documentation practices that reduce ambiguity and improve the reliability of LLM-assisted development.

\paragraph{4) Ambiguity Recognition Remains a Limiting Factor.}
Across all evaluated models, the precision of ambiguity detection is limited, leading to frequent false positives. This constrains the models' ability to reliably assess whether a requirement is ambiguous. As a result, developers should not assume that LLM outputs are trustworthy indicators of requirement clarity. Instead, ambiguous or complex inputs should be treated as potentially misinterpreted, and clarification strategies (e.g., prompting the model to restate requirements, explain assumptions, or outline its intended solution) should be employed to validate outputs before integration into codebases.

%% file: sections/7-related_work.tex
\subsection{Code Generation Benchmarks}

The evaluation of Large Language Models (LLMs) for code generation has evolved from controlled algorithmic tasks to more complex and realistic scenarios. Early benchmarks such as HumanEval~\cite{chen2021evaluating} and MBPP~\cite{austin2021program} consist of manually curated programming problems with concise, well-defined requirements and deterministic expected outputs. These benchmarks primarily focus on assessing functional correctness under unambiguous specifications.

To improve evaluation diversity and realism, recent benchmarks including MultiPL-E~\cite{cassano2023multipl}, BigCodeBench~\cite{zhuo2024bigcodebench}, LiveCodeBench~\cite{jain2024livecodebench}, and SWE-Bench~\cite{jimenez2023swe} extend the scope to multi-language settings, dynamic execution environments, and repository-level tasks. These efforts introduce more complex programming scenarios and better approximate real-world development conditions.

Despite these advances, a key characteristic shared by existing benchmarks is that their requirements are intentionally designed to be unambiguous and deterministic. While this design facilitates consistent evaluation and reproducibility, it abstracts away the uncertainty inherent in natural language specifications. As a result, these benchmarks primarily evaluate model performance under idealized conditions where each input corresponds to a single intended interpretation. They do not capture how models behave when requirements admit multiple valid interpretations. 

\subsection{Ambiguity in Requirements Engineering}

Ambiguity in Requirements Engineering refers to situations where a specification admits multiple valid interpretations~\cite{berry2004ambiguity,shah2015resolving}. It is recognized as a major source of defects, misunderstandings, and inconsistencies in software development. Prior research has extensively investigated ambiguity from a human-centric perspective, aiming to improve requirement quality and support stakeholders.

A range of techniques has been proposed for ambiguity detection and resolution. These include methods combining NLP with fuzzy set theory to identify ambiguities and clarify vague requirement terms~\cite{asadabadi2020ambiguous}, rule-based approaches for identifying structural ambiguities~\cite{toshiharu2022method}, and machine learning methods leveraging contextual representations such as BERT to address ambiguity~\cite{ezzini2022taphsir}. Additional work uses domain-specific word embeddings to identify cross-domain terminological ambiguity and knowledge graphs to detect pragmatic ambiguity~\cite{ferrari2019nlp,ferrari2012collective}.

These approaches are designed for human-in-the-loop scenarios, where ambiguity can be resolved through clarification, negotiation, and iterative refinement~\cite{fischbach2021practitioners}. However, LLM-based code generation operates under a different paradigm, in which the model must directly map a potentially ambiguous input to a single executable output without access to explicit clarification. This difference in operational assumptions limits the direct applicability of existing RE techniques and suggests that ambiguity should be re-examined in the context of automated code generation systems.

\subsection{Ambiguity-Aware Code Generation}

Recent work has begun to investigate how ambiguity affects LLM-based code generation. Existing efforts can be broadly categorized into benchmark-based evaluation and method-oriented approaches.

On the benchmark side, datasets such as AmbiQT~\cite{bhaskar2023benchmarking} and HumanEvalComm~\cite{wu2024humanevalcomm} incorporate ambiguous or underspecified requirements to evaluate model robustness. These benchmarks consider scenarios involving multiple valid solutions, vague instructions, and implicit intent inference. However, AmbiQT focuses on schema-dependent ambiguity in text-to-SQL generation, while HumanEvalComm evaluates whether models ask clarification questions when requirements contain ambiguity, inconsistency, or incompleteness. Neither provides paired analysis of multiple linguistic ambiguity types in general-purpose function-level code generation.

On the methodological side, several approaches address ambiguous or unclear specifications. Interactive frameworks such as ClarifyGPT~\cite{mu2024clarifygpt} and multi-agent systems~\cite{jia2025automated,fakhoury2024llm} introduce clarification mechanisms or collaborative reasoning processes to refine user intent. Other methods incorporate execution feedback, test cases, or iterative refinement strategies to improve code correctness.

Despite these efforts, it remains unclear how linguistic ambiguity types affect LLM behavior under matched task and test conditions. We address this gap by pairing each function-level task with four type-specific ambiguity variants and analyzing their effects on functional correctness, output consistency, and ambiguity handling.

%% file: sections/8-limitations.tex
Our evaluation focuses on function-level code generation, which enables controlled comparisons but leaves ambiguity in broader software engineering contexts, such as repository-level tasks with cross-function dependencies and evolving requirements, for future investigation. Each benchmark variant contains one ambiguity type introduced at a single location to isolate its effect. Future work can extend \bench to requirements containing multiple interacting ambiguities, incompleteness, or inconsistency.

%% file: sections/9-conclusion.tex
This paper studies how requirement ambiguity affects LLM-based code generation. We introduce \bench, a benchmark consisting of 1,304 function-level tasks with ambiguous requirements across four linguistic categories, and use it to systematically investigate model behavior under uncertain specifications. Our empirical results show that ambiguity consistently degrades generation performance and reduces functional consistency across model outputs. We further find that, although LLMs can often identify ambiguous requirements with relatively high recall, they exhibit limited precision and struggle to accurately localize and resolve the sources of ambiguity. Overall, our findings indicate that current LLM-based code generation systems are sensitive to requirement ambiguity and lack robustness in handling uncertain natural language specifications. These results highlight ambiguity as a critical factor in practical software development and motivate the need for ambiguity-aware approaches in future LLM-based software engineering systems.